\documentclass[conference]{IEEEtran}
\IEEEoverridecommandlockouts
\usepackage{cite}
\usepackage{amsmath,amssymb,amsfonts}
\usepackage{float}
\usepackage{subfigure}
\usepackage{algorithmic}
\usepackage{algorithm}
\usepackage{graphicx}
\usepackage{textcomp}
\usepackage{xcolor}
\usepackage[left=1.62cm,right=1.62cm,top=1.9cm]{geometry}
\usepackage{comment}
\usepackage{stackengine}
\def\BibTeX{{\rm B\kern-.05em{\sc i\kern-.025em b}\kern-.08em
    T\kern-.1667em\lower.7ex\hbox{E}\kern-.125emX}}
\begin{document}

\title{Learning Robust Representations for Communications over Interference-limited Channels
}

\author{
\IEEEauthorblockN{Shubham Paul \textsuperscript{*}, Sudharsan Senthil \textsuperscript{*}, Preethi Seshadri \textsuperscript{†}, Nambi Seshadri \textsuperscript{‡,*}, R David Koilpillai \textsuperscript{*}}
\IEEEauthorblockA{\textit{Email: paulshubham96@outlook.com, sudharsansenthil@hotmail.com, preethis@uci.edu,} \\{naseshadri@ucsd.edu, koilpillai@ee.iitm.ac.in} \\
*: Indian Institute of Technology Madras.
†: University of California Irvine.  ‡: University of California San Diego.
}
}\maketitle

\begin{abstract}
In the context of cellular networks, users located at the periphery of cells are particularly vulnerable to substantial interference from neighbouring cells, which can be represented as a two-user interference channel. This study introduces two highly effective methodologies, namely TwinNet and SiameseNet, using autoencoders, tailored for the design of encoders and decoders for block transmission and detection in interference-limited environments. The findings unambiguously illustrate that the developed models are capable of leveraging the interference structure to outperform traditional methods reliant on complete orthogonality. While it is recognized that systems employing coordinated transmissions and independent detection can offer greater capacity, the specific gains of data-driven models have not been thoroughly quantified or elucidated. This paper conducts an analysis to demonstrate the quantifiable advantages of such models in particular scenarios. Additionally, a comprehensive examination of the characteristics of codewords generated by these models is provided to offer a more intuitive comprehension of how these models achieve superior performance.
\end{abstract}


\section{Introduction}
In modern cellular systems (e.g., LTE and 5G), interference is a central issue because of the increased density of base stations and mobile users, all competing for limited spectrum resources. As user demand for high data rates and seamless connectivity grows, managing interference becomes even more critical. Users at the cell edges are particularly vulnerable because they are subject to strong interference from adjacent cells. This degrades their signal-to-interference-plus-noise ratio (SINR), leading to higher bit error rates (BER), increased latency and consequently low data rates. To address this, multi-cell networks employ advanced techniques like coordinated multi-point (CoMP)~\cite{CoMP}, where base stations coordinate their transmissions to reduce interference, and interference-aware resource scheduling \cite{interf-aware-WiMax}, which allocates spectrum dynamically to minimize conflicts. Massive MIMO \cite{massive-MIMO-interf} and beamforming also play a crucial role in mitigating interference by directing signal energy more precisely, reducing unintended interference to neighbouring cells. Despite these techniques, interference remains a significant challenge, as densification and spectrum reuse increase, requiring continuous innovation in interference management strategies to ensure high network performance.

Deep learning for the physical layer of communications involves leveraging advanced neural network architectures to improve communication systems' efficiency, performance, and adaptability. Some areas where deep learning may yield significant advantages include channel estimation, equalization, interference management, adaptive modulation and coding. The integration of deep learning into wireless communications uncovers a world of new opportunities. O'Shea explores this in~\cite{Tim_1}. The concept was extended and demonstrated for over-the-air systems in \cite{Dorner_1} and \cite{Dorner_2}. This ability of Deep learning models to address complex challenges and enhance the overall efficiency and performance of communication systems, warrants the use of such systems in the case of the two-user interference channel.  
The two-user interference channel represents a general communication scenario involving two transmitters and two receivers, where there is no cooperation between the transmitters and receivers. In this setup, each receiver decodes its intended message from a signal that is interfered with by the other user and affected by channel noise. Over time, there has been notable research focussing on communication in the presence of interference utilizing Deep Learning methodologies. Notably, Zhang introduced a deep autoencoder (DAE)-based communication scheme for the two-user Z-interference channel (ZIC) in \cite{Zhang}, further extended with finite-alphabet inputs in \cite{Zhang_1}. Fu proposes an intelligent successive interference cancellation (SIC) detection algorithm, namely I-SIC, for uplink non-orthogonal multiple access (NOMA) systems in \cite{Fu}. A novel deep learning-based near-orthogonal superposition (NOS) coding scheme is suggested by \cite{Bian} for the reliable transmission of short messages in the additive white Gaussian noise (AWGN) channel, catering to mMTC applications. Pellat introduces an innovative concurrent and interactive training method for a physical layer based on autoencoders in scenarios with multi-user interference channels in \cite{Pellatt}. Another major contribution was made in~\cite{Wu_2} where Wu addresses the dynamic interference in a multi-user Gaussian interference channel. However, to the best of our knowledge, it does not attempt to explain how the models actually "learned" to be robust in the presence of interference.

In this paper, an Autoencoder framework is presented for developing encoders and decoders tailored for block transmission and detection in an interference-limited environment. Two methodologies for constructing such models within this framework are outlined: TwinNet and SiameseNet. The findings demonstrate that the models outperform the perfectly orthogonal Time Division Multiple Access (TDMA) method in terms of Block Error Rate (BLER) performance. Additionally, a detailed analysis of the characteristics of the generated codewords is provided.

The rest of this paper is organized as follows: Sec~\ref{sec: System Model} formulates the problem and gives the system model. Sec~\ref{sec: Auto-encoder-based models} details the general Autoencoder-based framework, the two architectures, their training methodologies, and their operation. Sec.~\ref{sec: BLER Performance} shows the BLER performances of our models. And Sec.~\ref{sec: Latent Space Analysis} investigates the designed codewords and their properties. Our conclusions and possible future works are discussed in Sec~\ref{sec: conclusions}

\section{System Model}\label{sec: System Model}
In this paper, the performance of transmission and reception of blocks of bits is analyzed in the presence of an interfering user with Fully Connected Neural Network (FCNN) encoders and decoders within an autoencoder framework. 
A k-bit long message m is transmitted from a transmitter in a noisy channel across n symbols. If $n>k$, redundancy can be introduced in transmission, thereby yielding better BER and BLER performance. This behaviour essentially mimics block codes. A binary message is represented using One Hot Encoding (OHE), so a k-block size binary message becomes a $2^k$-length block vector M. The mappings from message $M_i$ of user-i to n-symbol codeword $z_i$ are learned by the FCNN block ~\eqref{eq: Encoder_eqn}. This can be viewed as a case of coded modulation. An average power constraint is imposed on the encodings produced to mimic the transmit power constraint~\eqref{eq: Power Cons}. The encoded vector $z_i \in \mathbb{R}^n$. Where, $n=k/ r$, is the size of the encoded dimension or the number of symbol transmissions per message block.

\begin{gather}
E_i \equiv F_{\theta_i}: M_i \rightarrow z_i \text{ ; i =1,2} 
\label{eq: Encoder_eqn}\\
\mathbb{E}\left[\lVert z^{2} \rVert]\right] = n
\label{eq: Power Cons}
\end{gather}

\begin{figure}[htbp]
\centerline{\includegraphics[width=0.9\linewidth]{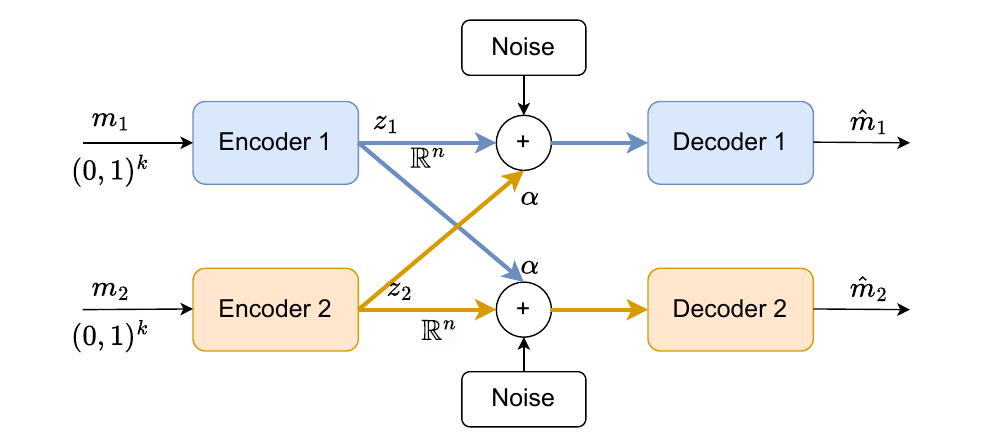}}
\caption{Two-user interference System Model}
\label{fig: System model 2}
\end{figure}

Consider a scenario where two users interfere with each other. Here, each user has a transmitter $Tx_i$, and a receiver, $Rx_i$ as depicted in Fig~\ref{fig: System model 2}. Let the transmitters $Tx_1$ and $Tx_2$ transmit $z_1$ and $z_2$ respectively. The received vectors $y_i$ at the $Rx_i$ is given by \eqref{eq:rx_eqn}, where $i\neq j$.

\begin{align}
    y_i = z_i + \alpha_j z_j + n_i \text{ ; } n_i \sim N(0, \sigma^2) \label{eq:rx_eqn}
\end{align}

Where $n_i$ denotes white noise samples and $\alpha_i$ is the interference strength of user-i's transmission at the $R_j$, here $j \neq i$. Throughout the paper, only symmetric interference is considered, hence $\alpha_1 = \alpha_2 = \alpha$. 
The task of the decoder $D_i$ is to find the best estimate $\hat{M_i}$ using the received vector $y_i$. An optimum decoder minimizes the probability of error in recovering the transmitted messages, given by \eqref{error_prob}.  

\begin{gather}
     D_i \equiv G_{\phi_i}:  y_i \rightarrow \hat{M_i}
     \text{ ; i=1,2}\label{eq: Dec_eq}\\
    P_{e} = \frac{1}{2^k}\sum_{j=1}^{j=2^{k}} P(\hat{M_j}\neq M_j) \label{error_prob}
\end{gather}
      
\section{Auto-encoder-based Models} \label{sec: Auto-encoder-based models}
Two autoencoder-based network architectures are introduced, namely, Twin Networks, and Siamese Networks, which can be employed for encoding and decoding data for one user in the presence of another interfering user. In the following sections, the performance of each network for varying interference strengths is analyzed and their respective latent spaces are investigated to understand their performance. 

\begin{figure}[htbp]
\centerline{\includegraphics[width=0.8\linewidth]{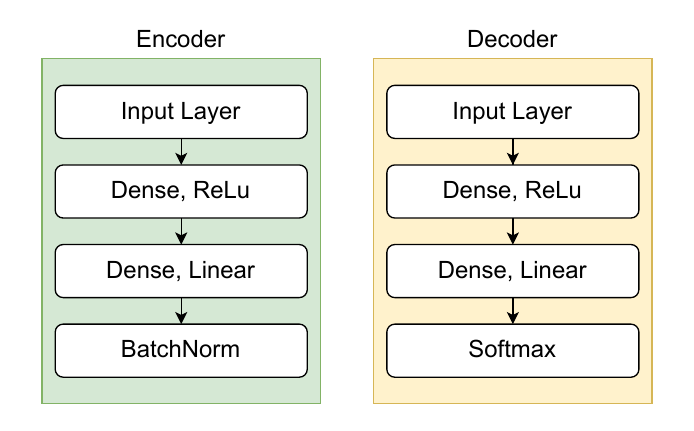}}
\caption{Encoder and Decoder Structure}
\label{fig: Encoder and Decoder Structure}
\end{figure}	
\subsection{General Framework}
\label{subsec: General Framework}
The parameters $\theta_i$ and $\phi_i$ of encoders and decoders respectively can be learned by optimizing the reconstruction loss of the respective users. The auto-encoder network consists of an encoder, followed by a non-trainable channel layer that distorts the latent space generated by the encoder and followed by a decoder that decodes the received vector. The Encoder and Decoder structure is given in Fig.~\ref{fig: Encoder and Decoder Structure}. The encoder has an initial input layer that accepts one-hot encoded representations of the input message. This is followed by a hidden Dense layer that is "\textit{ReLu}" activated, which in turn, is followed by a linearly activated layer and a "Batch-Norm" layer. A "Gaussian layer" is inserted to simulate the behaviour of the AWGN channel. The Decoder has one hidden layer which is "\textit{ReLu}" activated followed by a linear layer and an output layer which is "\textit{softmax}" activated. The interference from another user is introduced into the current user's channel as input.

Here, the channel is modelled as an AWGN channel whose variance (given by \eqref{eq: variance}) is set to satisfy the decided  $\frac{E_b}{N_0}$ for the system. To avoid confusion, subscripts to refer to specific users' messages are refrained from using, encodings, and received vector The decoder's final layer is a softmax layer, and $M^{*}$ is the output. $M^{*}[b], b \in [1,2^{k}]$ can be interpreted as the conditional probability that the transmitted message is $M_b$ on the reception of $y_i$, P($M=M_b | y_i$). $\hat{M}$ is estimated according to \eqref{eq:output_eqn}. 
\begin{gather}
    \sigma^2 = \frac{1}{2rE_b/N_0} \text{, } \label{eq: variance} \\
    \hat{M} = M_{\stackunder{argmax}{b} \text{ } M^{*}[b]}
    \label{eq:output_eqn}
\end{gather}

The network parameters $\phi_i \text{ and } \theta_i$ are optimized w.r.t the loss function. Specifically, the categorical cross-entropy. This loss function is based on the Kullback–Leibler divergence $D_{KL}$ between the true distribution $P_{M}(M_b)$ and the predicted distribution $Q_{M^{*}}(M_b)$ for one instance. Note that, $Q(M_b) = M^{*}[b]$. $H(P_{M}, Q_{M^{*}})$ is the categorical cross-entropy. In \eqref{eq:cat_cross_ent} only $H(P_{M}, Q_{M^{*}})$ depends on $\{\phi, \theta\}$ hence \eqref{eq:D_KL_min}.  
\begin{gather}
    D_{KL}(P_{M} \mid\mid Q_{M^{*}}) = \sum_b P_{M}(M_b) log(\frac{P_{M}(M_b)}{Q_{M^{*}}(M_b)})  \label{eq:KL_div}\\ 
   H(P_{M},Q_{M^{*}}) = - \sum_b P_{M}(M_b) log(Q_{M^{*}}(M_b))  \label{eq:cat_cross_ent}\\  
   \min_{\phi, \theta} D_{KL}(P_{M} \mid\mid Q_{M^{*}}) = \min_{\phi, \theta} H(P_{M},Q_{M^{*}}) \label{eq:D_KL_min} 
\end{gather}
 
\subsection{The Twin Networks (TwinNet) Architecture}
\label{subsec: Twin Networks}
\begin{figure}
  \centering
  {\includegraphics[width=0.9\columnwidth]{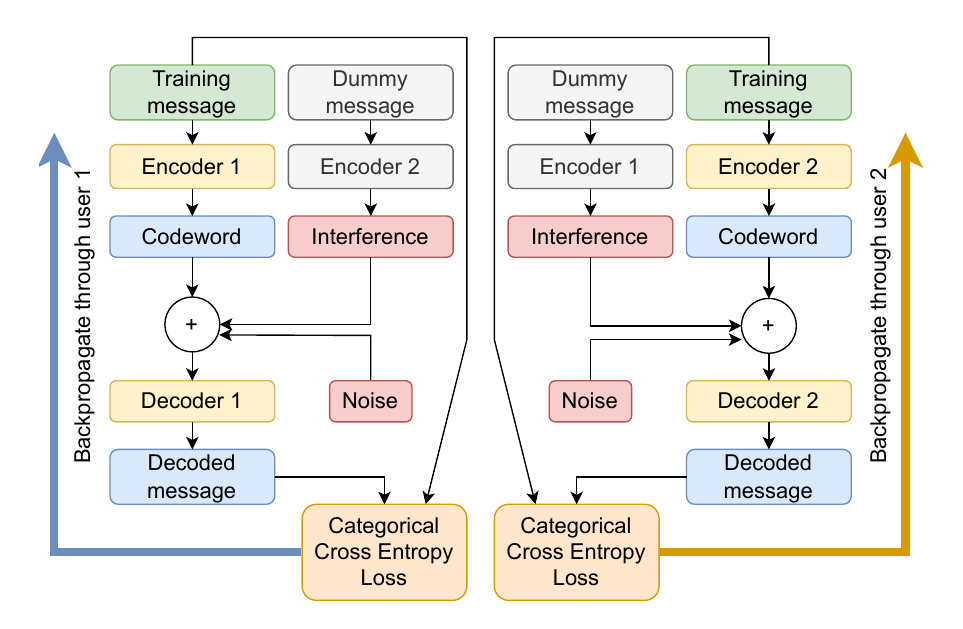}} 
  \caption{Twin Networks Architecture and Training.}
  \label{fig: Twin_net_architecture}
\end{figure}
 The transceiver architecture $Tx_i- Rx_i$ of the Twin-Networks is depicted in Fig.~\ref{fig: Twin_net_architecture}. Interference from the other user is given as an input while training the current user as explained in Algorithm~\ref{alg:twin-networks}. 

\textbf{Training and Operation: } In \cite{paper1} it is suggested that training with randomized SNR gives the best result for a single-user system, the same strategy is adapted to two-user systems. Referring to Algorithm~\ref{alg:twin-networks} $M_{i}$ is the one hot encoding of the message to be transmitted by user \textit{i} and $\hat{M_{i}}$ is the reconstructed one hot encoding of the message at the decoder of user \textit{i}. Interference is first generated from user-2 (step~\ref{alg:twin-networks interf gen}), which is then used to simulate interference at the user-1's transceiver line(step~\ref{alg:u2 interference}). The parameters of user-1, $\theta_1$ and $\phi_1$ are updated by optimizing its cost function $L[M^{*}_1,M_1]$ \eqref{eq:loss_eqn_twin}. Then, the same set of operations is carried out with the updated $E_{\theta_1}$ to simulate the interference to the user-2's transceiver line, and the parameters are updated by optimizing the corresponding cost. \eqref{alg: u2 loss_eqn_twin2} \\
\begin{align}
    L[\hat{M}_i, M^{*}_i]  = H(P_{M_i},Q_{M^{*}_i})
    \label{eq:loss_eqn_twin}
\end{align}

\begin{algorithm}
\caption{Algorithm to train the TwinNet model }\label{alg:twin-networks}

    \begin{algorithmic}[1]
           
\WHILE{ epoch $<$ maxepochs}

    \STATE $\frac{E_b}{N_0} \gets uniform[1,12]dB$; \\  
    \STATE $n_1 \gets AWGN((\sigma(\frac{E_b}{N_0}))$; $n_2 \gets AWGN((\sigma(\frac{E_b}{N_0}))$\\ 
    \STATE $intf_2 = E_{\theta_2}(M_2) \text{; } Z_1 = E_{\theta_1}(M_1)$ \\\label{alg:twin-networks interf gen}
    \STATE $y_1 = Z_1 + n_1 + intf2 $ \\ \label{alg:u2 interference}
    \STATE $M^{*}_1 = D_{\phi_1}(y_1)$ ; $C_1 = -L[M^{*}_1, M_1]$ \\ 
     \STATE $\{\theta_1\, \phi_1 \} \gets \{\theta_1, \phi_1,\}+ \{ \nabla_{\theta_1}C_1, \nabla_{\phi_1}C_1 \} $\\ \label{eq:alg_loss_eqn_twin}
     \STATE $intf_1 = E_{\theta_1}(M_1) \text{; } Z_2 = E_{\theta_2}(M_2)$\\
     \STATE $y_2 = Z_2 + n_2 + intf1 $ \\
     \STATE $M^{*}_2 = D_{\phi_2}(y_2)$ ; $C_2 = -L[M^{*}_i, M_2]$ \\
    \STATE $\{\theta_2, \phi_2 \} \gets \{\theta_2, \phi_2\}+ \{ \nabla_{\theta_2}C_2, \nabla_{\phi_2}C_2 \} $\\ \label{alg: u2 loss_eqn_twin2}
\ENDWHILE
\end{algorithmic}
\end{algorithm}
The TwinNet architecture doesn't allow the interfering user to learn from the error at the current user's end. As a result, the encoder and decoder of both users learn independently, treating interference from the other user similar to channel noise. This may lead to scenarios where the BLER performance of the two users is not similar.

\subsection{The Siamese Network (SiameseNet) Architecture}
\label{subsec: Siamese Network}
\begin{figure}
  \centering
  \includegraphics[width=0.9\columnwidth]{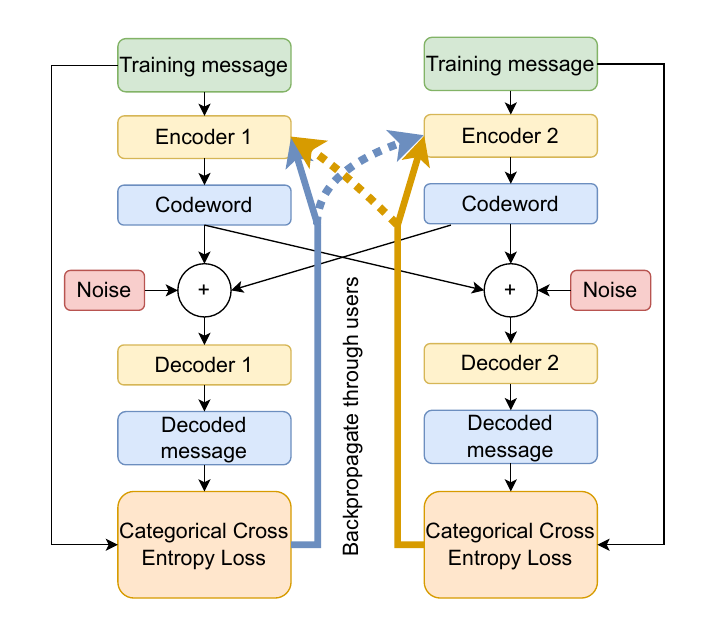} 
  \caption{Siamese Network Architecture and Training.}
  \label{fig:Siamese_arch}
\end{figure}
As discussed in the previous subsection, the performance of the two users in Twin networks need not be similar in terms of BLER as the networks are independently trained. The Siamese Network architecture is introduced to overcome the dissimilarity in performance of the two users. In Siamese Networks, the interfering user is allowed to learn based on the error incurred not only at its reconstruction, but also based on the error at the current user's reconstruction, and vice versa for the training of the current user. ~Fig. \ref{fig:Siamese_arch} shows the model architecture. A small perturbation in the parameters of $\theta_1$ will cause small changes $\triangle C_1$ and $\triangle C_2$ in $C_1$ and $C_2$ respectively and vice versa for $\theta_2$. These changes in the costs will be equally considered in updating the parameters $\theta_1 \text{, and } \theta_2$ enabling the encoders of both the users to learn jointly. Note that unlike $\theta_i$, a small perturbation in $\phi_1$ will not cause any change in $C_2$ same with $\phi_2$ hence the decoders of the two users are learned independently of each other. This is demonstrated in Fig.~\ref{fig:Siamese_arch} by the coloured arrows (denoting backpropagation) splitting at the encoders.

\textbf{Training and Operation:} Similar to Twin-Networks, Siamese Networks are trained in random SNR. Referring to Algorithm~\ref{alg:Siamese-network}, both $C_1$ and $C_2$ are considered to optimize each user's encoder parameters (steps~\ref{alg:Enc1_si_trn},~\ref{alg:Enc2_si_trn}), hence establishing cooperation in the encoding between the two users. Note that the decoder parameters are learned independently w.r.t their respective cost functions (step~\ref{alg:Dec_si_trn}). This is similar to the joint-encoding scheme, but after the training phase, both users can transmit independently of the other's transmission.\\

\begin{algorithm}
    \caption{Algorithm to train the Siamese network
    }\label{alg:Siamese-network}
     \begin{algorithmic}[1]   
\WHILE{ epoch $<$ maxepochs}

    \STATE $\frac{E_b}{N_0} \gets uniform[1,12]dB$; \\  
    \STATE $n_1 \gets AWGN((\sigma(\frac{E_b}{N_0}))$; $n_2 \gets AWGN((\sigma(\frac{E_b}{N_0}))$ \\  
    \STATE$z_1 = E_{\theta_1}(M_1) \text{; } z_2 = E_{\theta_2}(M_2) $ \\
    \STATE$y_1 = z_1 + z_2 + n_1 \text{; } y_2 = z_1 + z_2 + n_2 $ \\
    \STATE$M^{*}_1 = D_{\phi_1}(y_1) \text{; } M^{*}_2 = D_{\phi_2}(y_2)$ \\
    \STATE$C_1 = -L[M^{*}_1,M_1]$ ;  $C_2 = -L[M^{*}_2,M_2]$\\
    \STATE $ \{\phi_1,\phi_2 \} \gets \{ \phi_1, \phi_2 \} + \{ \nabla_{\phi_1}C_1, \nabla_{\phi_2}C_2 \} $ \\ \label{alg:Dec_si_trn}
   \STATE $\theta_1 \gets \theta_1 + \nabla_{\theta_1}C_1 + \nabla_{\theta_1}C_2$ \\ \label{alg:Enc1_si_trn}
   \STATE $\theta_2 \gets \theta_2 + \nabla_{\theta_2}C_1 + \nabla_{\theta_2}C_2$ \label{alg:Enc2_si_trn}
\ENDWHILE
\end{algorithmic}
\end{algorithm}

\section{BLER Performance}
\label{sec: BLER Performance}
\begin{figure}
    \centering
    \includegraphics[width = 0.8\columnwidth]{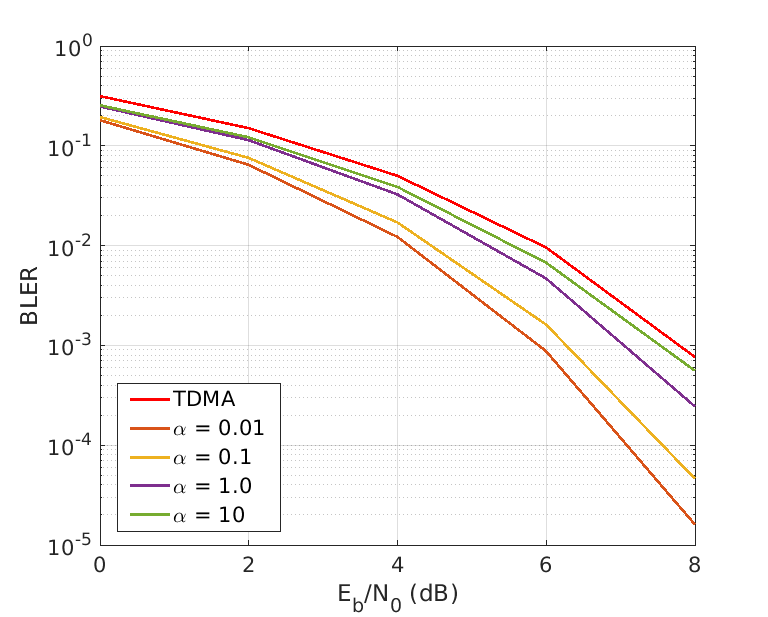}
    \caption{BLER Performance: TwinNet}
    \label{fig: BLER Performance TwinNet}

    \includegraphics[width = 0.8\columnwidth]{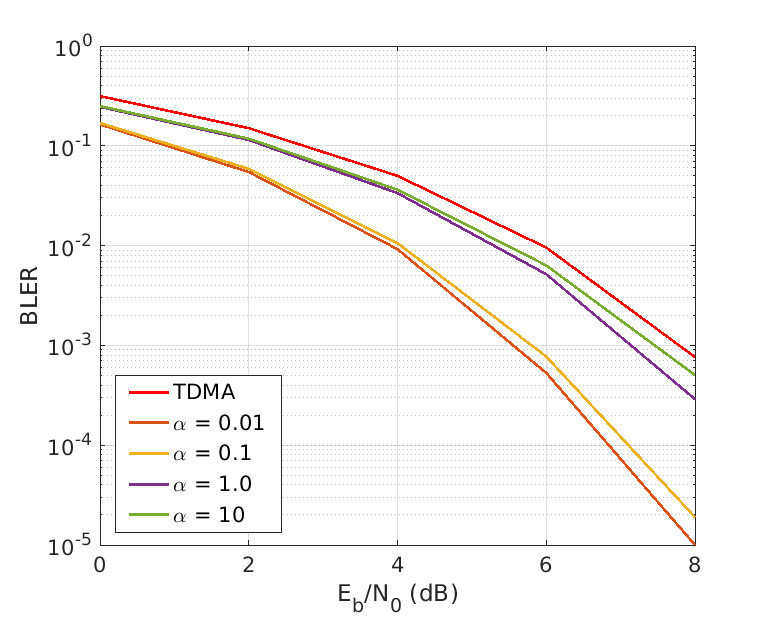}
    \caption{BLER Performance: SiameseNet}
    \label{fig: BLER Performance Siamese}
   
\end{figure}

This section studies the BLER performances of our models under different scenarios. k = 4 and n = 8 are considered and simulations are run for a range of $\frac{E_b}{N_0}$ values varying from 0 dB to 8 dB. The transmitted symbols from user 1 are allowed to interfere with symbols from user 2 as in \eqref{eq:rx_eqn}. We observe the results for interference strengths $\alpha=0.01, 0.1, 1\text{ and }10$. To ensure a fair comparison in spectral efficiency, we compare it to the TDMA case where the two users are separated in time and send 4 uncoded BPSK symbols per block. Fig.~\ref{fig: BLER Performance TwinNet} shows the BLER performance of the Twin-Net models and Fig.~\ref{fig: BLER Performance Siamese} shows the same for the SiameseNet models. Our results demonstrate that even at high interference, both models are resilient and yield performances as good as being perfectly orthogonal. Even more interesting is the observation that for moderate interferences ($\alpha = 1$), we observe a 1dB gain when compared to orthogonal transmissions. This gain only improves as the strength of interference reduces. This can be attributed to the redundancy introduced into the transmitted codewords. 

In further experimentation, the models are exclusively trained for a specific value of $\alpha$, and subsequently, their performance across scenarios featuring varying values of $\alpha$ is assessed. Specifically, a model is trained for $\alpha =10$ and is evaluated for cases $\alpha =  1, 10 \text{ and } 20$. An interesting observation is that, despite a doubling in interference strength for $\alpha =20$, the SiameseNet model maintained a comparable BLER performance. However, for $\alpha = 1$, the model under-performed than the model trained on $\alpha = 1$. Conversely, the TwinNet model did not demonstrate equivalent generalization capabilities. Specifically, for $\alpha =20$, the model significantly underperformed the Siamese model. However, for $\alpha = 1$, although some loss was evident, it remained minimal. This indicates that the codewords from SiameseNet are more resilient to interference than TwinNet for $\alpha =10$. More discussions on the codewords are given in Sec~\ref{sec: Latent Space Analysis}.

\begin{figure}[htbp]
  \centering
  \includegraphics[width=0.8\linewidth]{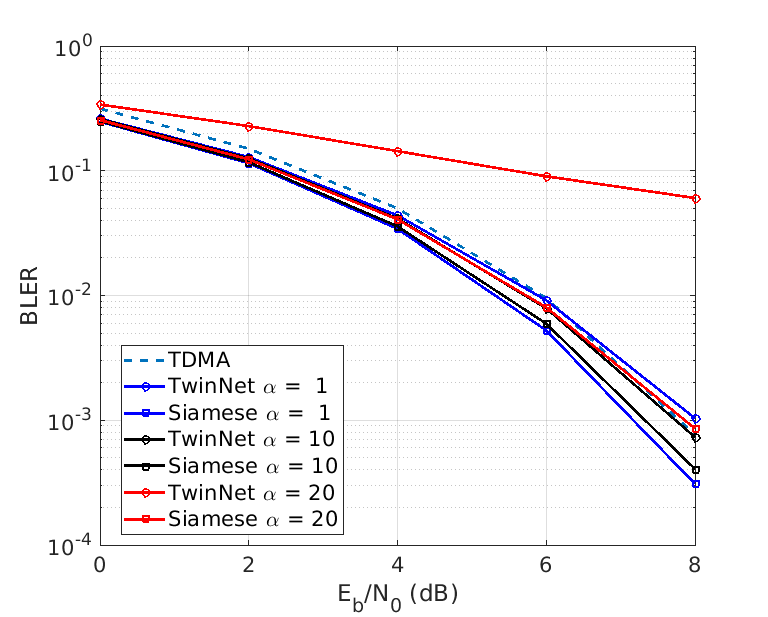} 
  \caption{BLER Performance of $\alpha =10$ model}
  \label{fig: BLER Performance varying interference}
\end{figure}

\section{Latent Space Analysis}
\label{sec: Latent Space Analysis}

This section investigates the latent space encodings, i.e. the codewords created by the Encoders of our models. The objective is to explain the BLER performances previously examined by investigating two key properties of the resulting codewords: the pairwise distances between codewords and their pairwise correlations. The self and cross distances can be determined using Eqn.~\eqref{eq: distances} while the correlations can be calculated using Eqn.~\eqref{eq: correlations}. 

\begin{subequations}
\label{eq: distances} 
\begin{gather}
    d_{self} = ||z_{k,i} - z_{k,j}||,\text{ where }i \neq j \\    
    d_{cross}= ||z_{k,i} - z_{l,j}||, \text{ where } k\neq l 
\end{gather}
\end{subequations}

\begin{subequations}
\label{eq: correlations}
\begin{gather}
    R_{self} = <z_{k,i} \cdot z_{k,j}>,\text{ where }i \neq j \\    
    R_{cross}= <z_{k,i} \cdot  z_{l,j}>, \text{ where } k\neq l 
\end{gather}
\end{subequations}

The error performance of the trained models is influenced by the distances between the codewords. Greater separation between codewords in the latent space reduces the likelihood of decoding errors, highlighting the importance of studying mutual distances. Table~\ref{tab: Pairwise Distances} provides a comprehensive list of these distances. In scenarios where the system is noise-limited, i.e. for $\alpha<1$, self-distances ($d_{self}$) have a higher impact than cross-distances ($d_{cross}$), leading the models to prioritize maximizing the former. Conversely, in interference-limited environments, the models focus on maximizing cross-distances. However, this analysis does not fully elucidate the model's behaviour under high interference conditions. 
\begin{table}
    \centering
    \begin{tabular}{|c|c|c|c|c|} \hline 
         &  $\alpha = $0.01&  $\alpha = $0.1&  $\alpha = $1& $\alpha = $10\\ \hline 
 Min $d_{cross}$ TwinNet& 1.35& 1.58& 3.89&3.85\\ \hline 
         Min $d_{self}$ TwinNet&  3.34&  3.28&  2.98& 3.07\\\hline
 Min $d_{cross}$ SiameseNet& 1.1& 0.79& 3.91&3.98\\\hline
 Min $d_{self}$ SiameseNet& 3.54& 3.58& 2.97&2.89\\\hline
    \end{tabular}
    \caption{Pairwise Distances}
    \label{tab: Pairwise Distances}
\end{table}

\begin{figure}
    \centering
    \includegraphics[width=0.9\linewidth]{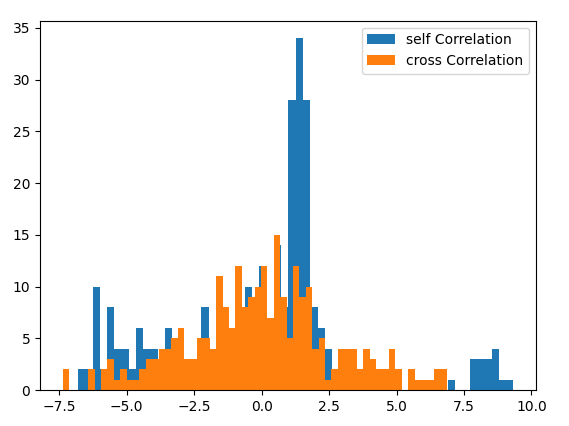}
    \caption{Correlation for $\alpha=0.1$ TwinNet}
    \label{fig: Corr 0p1 TwinNet}
    \includegraphics[width=0.9\linewidth]{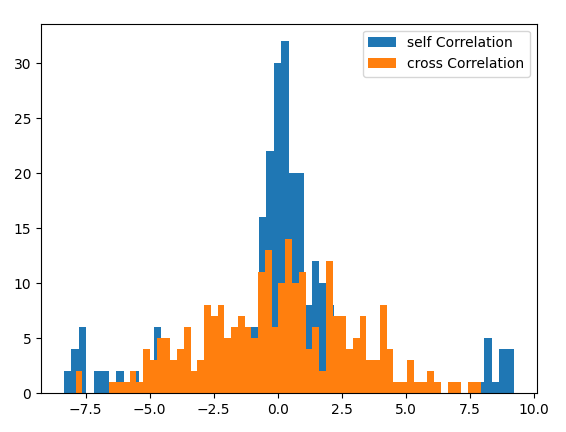}
    \caption{Correlation for $\alpha=0.1$ SiameseNet}
    \label{fig: Corr 0p1 SiameseNet}
\end{figure}

\begin{figure}
    \centering
    \includegraphics[width=0.9\linewidth]{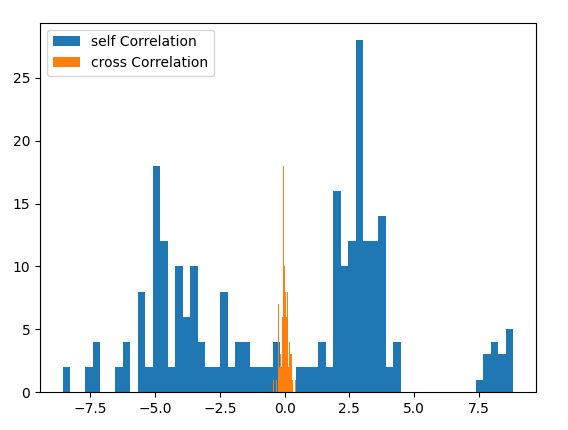}
    \caption{Correlation for $\alpha=10$ TwinNet}
    \label{fig: Corr 10 TwinNet}
    \includegraphics[width=0.9\linewidth]{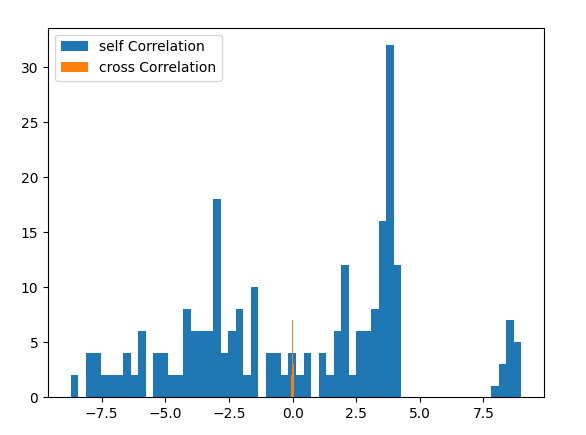}
    \caption{Correlation for $\alpha=10$ SiameseNet}
    \label{fig: Corr 10 SiameseNet}
\end{figure}

In Sec~\ref{sec: BLER Performance}, it was observed that both TwinNet and Siamese demonstrated similar performance when the evaluation data's interference value matched that of the training data. However, an interesting property was noted in the SiameseNet model. Specifically, when the model was trained at a certain interference value, it showed the ability to generalize well for other interference values. This property is more prominent for $\alpha>10$. Despite this observation, the study of pairwise distances did not fully explain this behaviour. Subsequently, an investigation into the correlation between the codewords was conducted. The analysis, depicted in Fig.~\ref{fig: Corr 10 TwinNet} and Fig.~\ref{fig: Corr 10 SiameseNet} for $\alpha =10$, the cross-correlation tends to 0, i.e. the set of codewords becomes uncorrelated to each other. In the case of SiameseNet, the correlation range was -0.075 to 0.064, while for TwinNet, it varied from -0.428 to 0.443. This suggests that the codewords produced by SiameseNet are less coupled to the interferer, which justifies its superior performance compared to TwinNet when the training and evaluation interferences differ. Furthermore, for smaller interference values, as indicated by Fig.~\ref{fig: Corr 0p1 TwinNet} and Fig.~\ref{fig: Corr 0p1 SiameseNet}, the system is noise-limited and the impact of the interferer, though not negligible, is less significant. Thus, the values of $R_{cross}$ are more diverse and the focus is to learn codewords such that tend to 0 which leads to a higher number of codeword pairs with $R_{self} = 0$. Here again, the SiameseNet outperforms the TwinNet model.

\section{Conclusions}
\label{sec: conclusions}
In this research paper, we have introduced the TwinNet and SiameseNet methodologies for block transmission and detection in an interference-limited environment. Our study demonstrates that the BLER performance of our models surpasses that of the perfectly orthogonal TDMA method. We have conducted a comprehensive analysis of the characteristics of the codewords generated by our models. In high interference scenarios, the models have been able to effectively separate themselves from the interferer while maintaining satisfactory performance. Notably, the SiameseNet exhibits significantly more resilience in BLER performance in high interference scenarios, more so when the interference strength is substantially more than what the model was trained for.  An attribute of its ability to learn the codewords jointly. Our future research will delve deeper into the information-theoretic analysis of the codeword space and evaluation of the performance against Shannon capacity. Future works will also study multiple code rates and observe the model performances and the resultant codewords.

\bibliographystyle{IEEEtran}
\bibliography{main.bib}

\begin{thebibliography}{10}
\providecommand{\url}[1]{#1}
\csname url@samestyle\endcsname
\providecommand{\newblock}{\relax}
\providecommand{\bibinfo}[2]{#2}
\providecommand{\BIBentrySTDinterwordspacing}{\spaceskip=0pt\relax}
\providecommand{\BIBentryALTinterwordstretchfactor}{4}
\providecommand{\BIBentryALTinterwordspacing}{\spaceskip=\fontdimen2\font plus
\BIBentryALTinterwordstretchfactor\fontdimen3\font minus \fontdimen4\font\relax}
\providecommand{\BIBforeignlanguage}[2]{{%
\expandafter\ifx\csname l@#1\endcsname\relax
\typeout{** WARNING: IEEEtran.bst: No hyphenation pattern has been}%
\typeout{** loaded for the language `#1'. Using the pattern for}%
\typeout{** the default language instead.}%
\else
\language=\csname l@#1\endcsname
\fi
#2}}
\providecommand{\BIBdecl}{\relax}
\BIBdecl

\bibitem{CoMP}
S.-h. Xiao and Z.-p. Zhang, ``Coordinated multipoint transmission systems with the clustered super-cell structure configuration,'' in \emph{2009 5th International Conference on Wireless Communications, Networking and Mobile Computing}, 2009, pp. 1--4.

\bibitem{interf-aware-WiMax}
H.-Y. Wei, S.~Ganguly, R.~Izmailov, and Z.~Haas, ``Interference-aware ieee 802.16 wimax mesh networks,'' in \emph{2005 IEEE 61st Vehicular Technology Conference}, vol.~5, 2005, pp. 3102--3106 Vol. 5.

\bibitem{massive-MIMO-interf}
C.~Lee, C.-B. Chae, T.~Kim, S.~Choi, and J.~Lee, ``Network massive mimo for cell-boundary users: From a precoding normalization perspective,'' in \emph{2012 IEEE Globecom Workshops}, 2012, pp. 233--237.

\bibitem{Tim_1}
T.~O’Shea and J.~Hoydis, ``An introduction to deep learning for the physical layer,'' \emph{IEEE Transactions on Cognitive Communications and Networking}, vol.~3, no.~4, pp. 563--575, 2017.

\bibitem{Dorner_1}
S.~Dorner, S.~Cammerer, J.~Hoydis, and S.~ten Brink, ``On deep learning-based communication over the air,'' in \emph{2017 51st Asilomar Conference on Signals, Systems, and Computers}, 2017, pp. 1791--1795.

\bibitem{Dorner_2}
S.~Dörner, S.~Cammerer, J.~Hoydis, and S.~t. Brink, ``Deep learning based communication over the air,'' \emph{IEEE Journal of Selected Topics in Signal Processing}, vol.~12, no.~1, pp. 132--143, 2018.

\bibitem{Zhang}
X.~Zhang and M.~Vaezi, ``Deep autoencoder-based z-interference channels,'' in \emph{2023 IEEE Wireless Communications and Networking Conference (WCNC)}, 2023, pp. 1--6.

\bibitem{Zhang_1}
X.~Zhang, M.~Vaezi, and L.~Zheng, ``Interference-aware constellation design for z-interference channels with imperfect csi,'' in \emph{ICC 2023 - IEEE International Conference on Communications}, 2023, pp. 6385--6390.

\bibitem{Fu}
J.~Fu, Y.~Xiao, H.~Liu, P.~Yang, and B.~Zhang, ``A novel intelligent sic detector for noma systems based on deep learning,'' in \emph{2021 IEEE 93rd Vehicular Technology Conference (VTC2021-Spring)}, 2021, pp. 1--6.

\bibitem{Bian}
C.~Bian, M.~Yang, C.-W. Hsu, and H.-S. Kim, ``Deep learning based near-orthogonal superposition code for short message transmission,'' in \emph{ICC 2022 - IEEE International Conference on Communications}, 2022, pp. 3892--3897.

\bibitem{Pellatt}
L.~Pellatt, M.~Nekovee, and D.~Wu, ``A concurrent training method of deep-learning autoencoders in a multi-user interference channel,'' in \emph{2021 17th International Symposium on Wireless Communication Systems (ISWCS)}, 2021, pp. 1--6.

\bibitem{Wu_2}
D.~Wu, M.~Nekovee, and Y.~Wang, ``Deep learning-based autoencoder for m-user wireless interference channel physical layer design,'' \emph{IEEE Access}, vol.~8, pp. 174\,679--174\,691, 2020.

\bibitem{paper1}
\BIBentryALTinterwordspacing
S.~Senthil, S.~Paul, N.~Seshadri, and R.~D. Koilpillai, ``Learning robust representations for communications over noisy channels,'' 2024. [Online]. Available: \url{https://arxiv.org/abs/2409.01129}
\BIBentrySTDinterwordspacing

\end{thebibliography}

\end{document}